\documentclass[prl,superscriptaddress,twocolumn,showpacs,preprintnumbers,amsmath,amssymb]{revtex4}

\usepackage{graphicx}
\usepackage{bm}
\usepackage{color}

\def\tons{\ensuremath{T^{\rm ons}}}
\def\tc{\ensuremath{T_{\rm c}}}
\def\tstar{\ensuremath{T^*}}

\def\sbl{\ensuremath{\sigma_{\rm bl}}}
\def\sint{\ensuremath{\sigma_{\rm int}}}
\def\cm{\ensuremath{\rm cm^{-1}}}
\def\etal{{\it et~al.,}}

\def\sonebl{\ensuremath{\sigma_{1,{\rm bl}}}}
\def\soneint{\ensuremath{\sigma_{1,{\rm int}}}}

\begin{document}

\preprint{???}

\title{Evidence of precursor superconductivity as high as 180 K from infrared spectroscopy}

\author{A. Dubroka}
\author{M. R\"{o}ssle}
\author{K. W. Kim}
\author{V. K. Malik}
 \affiliation{University of Fribourg, Department of Physics and Fribourg Center for Nanomaterials,
Chemin du Mus\'{e}e 3, CH-1700 Fribourg, Switzerland}
\author{D. Munzar}
\affiliation{Department of Condensed Matter Physics, Faculty of Science, Masaryk University, Kotl\'a\v{r}sk\'a 2, 61137 Brno, Czech Republic}
\author{D. N. Basov}
\author{A. Schafgans}
\author{S. J. Moon}
\affiliation{Department of Physics, University of California, San Diego, La Jolla, California 92093, USA}
\author{C.~T.~Lin}
\author{D. Haug}
\author{V. Hinkov}
\author{B. Keimer}
\affiliation{Max-Planck-Institut f\"ur Festk{\"o}rperforschung, Heisenbergstrasse 1, D-70569 Stuttgart, Germany}
\author{Th. Wolf}
\affiliation{Karlsruhe Institute of Technology, Institut f\"ur Festk{\"o}rperphysik, D-76021 Karlsruhe, Germany}
\author{J. G. Storey}
\author{J. L. Tallon}
\affiliation{MacDiarmid Institute, Industrial Research Ltd., P.O. Box 31310, Lower Hutt, New Zealand}
\author{C. Bernhard}%
 \email{christian.bernhard@unifr.ch}
\affiliation{University of Fribourg, Department of Physics and Fribourg Center for Nanomaterials,
Chemin du Mus\'{e}e 3, CH-1700 Fribourg, Switzerland}

\date{\today}

\begin{abstract}
We show that a multilayer analysis of the infrared  $c$-axis response of RBa$_2$Cu$_3$O$_{7-\delta}$ (R=Y, Gd, Eu) provides important new information about the anomalous normal state properties of underdoped cuprate high temperature superconductors. Besides competing correlations which give rise to a pseudogap that depletes the low-energy electronic states below $\tstar\gg\tc$, it enables us to identify the onset of a precursor superconducting state below $\tons>\tc$. We map out the doping phase diagram of \tons\  which reaches a maximum of ~180~K at strong underdoping and present magnetic field dependent data which confirm our conclusions.

\end{abstract}

\maketitle

The anomalous normal state properties of underdoped cuprate high temperature superconductors (HTSC) and, in particular, the so-called pseudogap phenomenon remain the subject of an intense debate~\cite{Timusk99,Hufner08,Basov05}. The wide spectrum of interpretations ranges from a precursor superconducting state (PSC) to electronic correlations that compete with superconductivity (SC). The conflicting experimental data may even be explained in terms of a dual scenario where a PSC develops at $\tons>\tc$ in the presence of a competing pseudogap that depletes the low-energy electronic states below $\tstar\gg\tons$. However, it remains a challenging experimental task to identify these transitions and to disentangle their contributions to the electronic response. In the following we show that this goal can be achieved based on a multilayer analysis of the infrared $c$-axis conductivity of RBa$_2$Cu$_3$O$_{7-d}$ (R=Y, Gd, Eu) single crystals.

\begin{figure*}
\vspace{-0.7cm}
\hspace*{-0.8cm}\includegraphics[width=15cm]{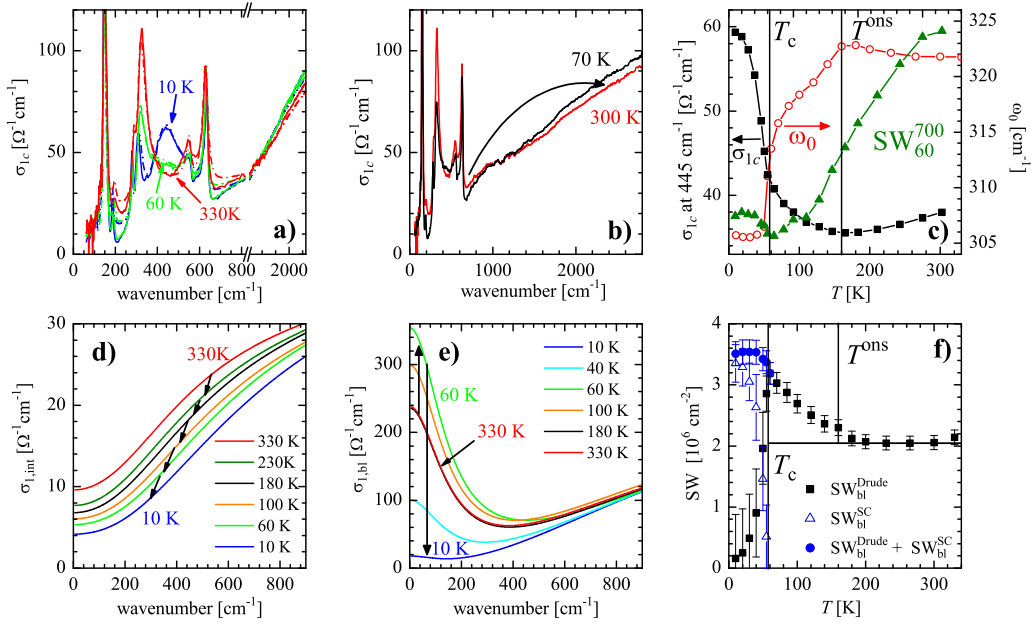} 
\vspace{-0.8cm}
\caption{\label{MLM} Multilayer-model analysis of the $c$-axis response of Y-123 with $\tc=58$~K and its local conductivities. (a) Measured (solid lines) and modeled (dash-dotted lines) real part of the conductivity, $\sigma_{1c}$. (b) Expanded scale showing the upwards shift of the spectral weight below $\tstar>300$~K. (c) Temperature dependence of the phonon frequency, $\omega_0$ (open circles), conductivity at the maximum of the TPM (solid squares), and the SW between 60 and 700~\cm\ (solid triangles). (d), (e), Real part of the local conductivities of the inter-bilayer region, \soneint, and the intra-bilayer one, \sonebl, respectively. (f), Temperature dependence of the spectral weight of the Drude response, SW$_{\rm bl}^{\rm Drude}$ and the delta-function response below \tc, SW$_{\rm bl}^{\rm SC}$.
}
\end{figure*}

Studies of the infrared $c$-axis conductivity, $\sigma_c(\omega)=\sigma_{1c}(\omega)+{\rm i}\sigma_{2c}(\omega)$, of underdoped YBa$_2$Cu$_3$O$_{7-\delta}$ (Y-123) where the first to show that besides the so-called spin gap~\cite{Alloul89}, a partial, gap-like suppression also occurs in the low-energy charge excitations~\cite{Homes93}. Recent measurements demonstrated that this pseudogap competes with SC, since it removes low-energy spectral weight which is shifted into a band above the gap edge~\cite{Li08}. This is unlike the SC gap at $T<\tc$ where the missing spectral weight is redistributed into a delta function at $\omega=0$ that accounts for the loss-free response of the SC condensate. The suppression of $\sigma_{1c}$ at $T>\tc$ thus provides direct information about the temperature- and energy scales, $\tstar$ and $\Delta^{\rm PG}$, of the competing pseudogap.
Here we outline that the $c$-axis response of underdoped Y-123 also contains clear signatures of a PSC state. They are contained in the so-called transverse plasma mode (TPM) and the anomalous temperature dependence of certain infrared-active phonons which both become most pronounced below $\tc$ but gradually develop already at $\tons>\tc$~\cite{Homes95,Schutzmann00, Bernhard00,Marel96}.  The understanding of these features and their relationship to the PSC require consideration of the layered structure of Y-123 which contains two CuO$_2$ planes per unit cell (so-called bilayer unit). A remarkable forte of infrared spectroscopy is in the ability to probe the local conductivity inside the bilayer units and between these units, \sbl\ and \sint, respectively. A quantitative description of both TPM and the related phonon anomalies is provided by the so-called multilayer model (MLM) of the $c$-axis electrodynamics~\cite{Marel96,Munzar99,Gruninger00}.

RBa$_2$Cu$_3$O$_{7-\delta}$ (R=Y, Gd, Eu) crystals of typical dimension 2$\times$2$\times$0.5-1 mm$^3$ were grown in Y-stabilized zirconium crucibles as described in Ref.~\cite{Schlachter00}. The hole doping of the CuO$_2$ planes, $p$, was adjusted via the oxygen content of the CuO chain layer, $\delta$, by annealing in flowing O$_2$ and subsequent rapid quenching in liquid nitrogen. Some strongly underdoped samples were also obtained by partial substitution of R$^{3+}$ with Ca$^{2+}$~\cite{Tallon95}. The quoted \tc\  values were determined by dc magnetization measurements. The $p$ values were obtained either with the empirical relationship~\cite{Tallon95}, from the measured thermo-electric power (TEP) at room temperature~\cite{Tallon95}, or from the Ca-content according to $p=x/2$.
The ellipsometric measurements were performed with a home-built ellipsometer attached to a Bruker Fast-Fourier spectrometer below 700~\cm\ at the infrared-beamline of the ANKA synchrotron at KIT Karlsruhe, Germany and at 400-4000~\cm\ with a similar lab-based setup~\cite{Bernhard04}. The presented $c$-axis polarized spectra are corrected for anisotropy effects by using standard numerical procedures. 
The magnetic field dependence of the $c$-axis response was measured with a near-normal-incidence reflection and in-situ gold evaporation technique using the split-coil superconducting magnet at UCSD as described in Ref.~\cite{LaForge07}. 

\begin{figure*}[t!]
\vspace{-0.7cm}
\hspace*{-0.6cm}
\includegraphics[width=19cm]{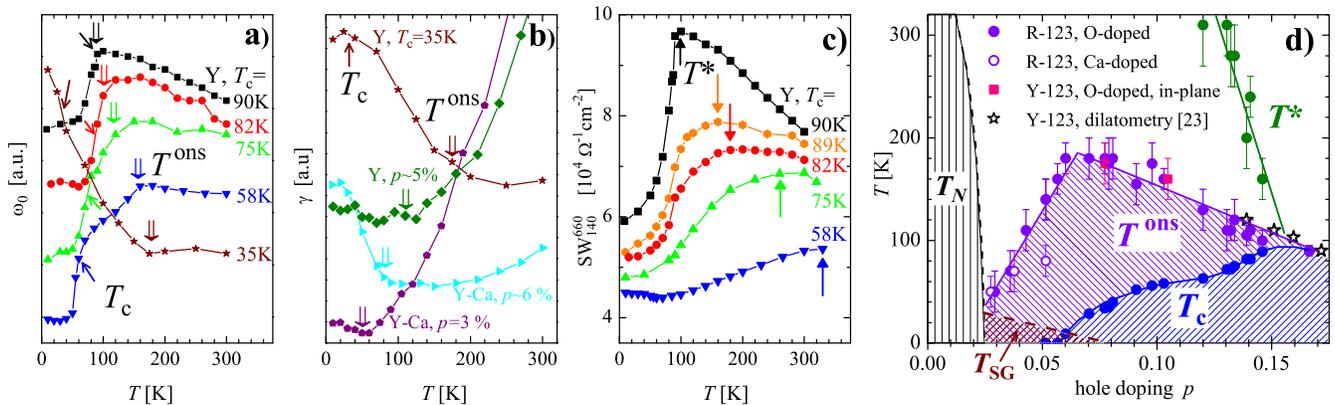} 
\vspace{-1.2cm}
\caption{\label{phdiag} (a) Temperature dependence of the frequency of the 320 \cm\ phonon for $p>0.08$ shown on a renormalized scale. The onset of the anomalous softening at \tons\ and \tc\ is marked by the thick and thin arrows, respectively. (b) Corresponding changes for samples with $p\leq0.08$ where \tons\ is obtained from the anomalous broadening of the phonon. (c) Temperature dependence of the spectral weight between 140 and 660~\cm\ 
(SW$_{140}^{660}$) where \tstar\ is determined from the onset of an anomalous decrease as marked by the arrows. (d) Resulting doping phase diagram of \tstar, \tons\ and \tc.}
\end{figure*}

Figure~\ref{MLM} shows the results of the MLM analysis for underdoped YBa$_2$Cu$_3$O$_{6.6}$ with $\tc=58$ (2)K, for details see~Ref.~\cite{SOM} part I. Figures~\ref{MLM}(a) and~\ref{MLM}(b) compare the experimental and fitted spectra of $\sigma_{1c}(\omega)$ and demonstrate that the MLM provides an excellent account of the features that are relevant to the forthcoming discussion, such as 
(i) the TPM near 450~\cm, (ii) the anomaly of the 320~\cm\ phonon, and (iii) the suppression of the electronic background by the competing pseudogap with $2\Delta^{\rm PG}\approx1200\ \cm$. Their temperature dependence is detailed in Fig.~\ref{MLM}(c) in terms of (i) $\sigma_{1c}$(445~\cm), (ii) the phonon frequency, $\omega_0$, and (iii) the spectral weight between 60 and 700~\cm, SW$_{60}^{700}$. It confirms that the competing pseudogap (iii) appears at a significantly higher temperature of $\tstar>300~$K than the features (i) and (ii) which develop concurrently below $\tons\approx160$~K. Figures~\ref{MLM}(d) and~\ref{MLM}(e) show the real parts of the local conductivities, \sint\ and \sbl, as obtained with the MLM. The low values and the insulator-like temperature and frequency dependence of $\sigma_{1,\rm int}$ are characteristic of incoherent transport and reflect the competing pseudogap below $\tstar>300$~K. 
The important, new information is contained in \sonebl\ which contains a sizeable Drude-like component that is characteristic of a coherent response. Its temperature dependence is detailed in Fig.~\ref{MLM}(f) in terms of the spectral weight of the Drude-component, SW$_{\rm bl}^{\rm Drude}$, and of the delta function at $\omega=0$, SW$_{\rm bl}^{\rm SC}$. The latter represents the macroscopically-coherent condensate below \tc\ and has been deduced from $\sigma_{2,\rm bl}$ (not shown). Notably, SW$_{\rm bl}^{\rm Drude}$ is not affected at $\tstar$ and remains constant down to $\tons\approx160$~K below which it starts to increase. Below \tc, the delta function develops and acquires most of the spectral weight. We emphasize that these trends are as expected for a PSC. The increase of SW$_{\rm bl}^{\rm Drude}$ below $\tons\approx160$~K is caused by the formation of the precursor SC gap which, similar to the SC gap below \tc, causes a downward shift of SW towards zero frequency. The response of the partially coherent condensate at $\tc<T<\tons$ shows up as a Drude-like peak whose width reflects the finite superconducting correlation time. Below \tc, as these fluctuations are suppressed, all of this spectral weight is finally transferred to the delta function. Note that a markedly different behavior would occur for a spin- or charge density wave state where a significant part of the low-frequency spectral weight would be removed and shifted to higher frequencies.

These observations raise the question why the signatures of the PSC appear predominantly in \sonebl\ while that of the competing pseudogap appears mainly in \soneint. This is at least partially due to the inter-bilayer and intra-bilayer hopping matrix elements, 
$t_{\perp,\rm int}({\mathbf k}_{\parallel})$ and  $t_{\perp,\rm bl}({\mathbf k}_{\parallel})$, respectively, and their dependence on the in-plane wave vector, ${\mathbf k}_\parallel$. While $t_{\perp,\rm int}({\mathbf k}_{\parallel})$ is maximal at the boundary of the Brillouin zone (antinodal region) and vanishes along the Brillouin zone diagonal (nodal region), $t_{\perp,\rm bl}({\mathbf k}_{\parallel})$ depends only weakly on ${\mathbf k}_\parallel$ (for details see~Ref.~\cite{SOM} part I). Since the pseudogap is localized around the antinodal region leaving ungapped Fermi arcs around the nodal region~\cite{Tanaka06}, \sint\ is dominated by the pseudogap while \sbl\ contains a major contribution from the nodal quasiparticles. This interpretation is confirmed by our finding that the in-plane response exhibits qualitatively similar changes below \tons\ as the ones in \sbl\ (see~Ref.~\cite{SOM} part III).

The scenario of a PSC is furthermore supported by the doping dependence of \tons\ of a series of R-123 crystals (R=Y, Eu, Gd). The value of \tons\ has been conveniently deduced from the anomalous temperature dependence of the 320 \cm\ phonon mode that coincides with that of the TPM [see Fig.~\ref{MLM}(c) and Ref.~\cite{Bernhard00,Munzar99}]. For samples with $p>0.08$, we employed the anomalous phonon softening [Fig.~\ref{phdiag}(a)], while at $p\leq0.08$, where the TPM merges with the 320 \cm\ phonon, we focused on the anomaly in the linewidth, $\gamma$ [Fig.~\ref{phdiag}(b)]. The corresponding \tstar\ values have been derived from the suppression of SW$_{140}^{660}$ [Fig.~\ref{phdiag}(c)]. The resulting doping dependencies of \tc, \tons, and \tstar\ are displayed in Fig.~\ref{phdiag}(d). It highlights that \tons\ emerges from the \tc\ line of the overdoped samples and increases on the underdoped side until it reaches a maximum of $\tons\approx180$~K close to the boundary of static antiferromagnetism~\cite{Niedermayer98,Sanna04}. Subsequently, \tons\ decreases sharply as the metal-to-insulator transition is further approached. The characteristic evolution of \tons, in particular its dome-like shape, is consistent with the assignment to the PCS. It also agrees with previous reports based on Nernst-effect and magnetization~\cite{Wang06,Li10}, thermal expansion~\cite{Meingast01}, specific heat~\cite{Tallon10}, and recent scanning-tunneling spectroscopy measurements~\cite{Lee09}.
We emphasize that a fairly large amount of spectral weight is involved in the changes at $\tc<T<\tons$. The fraction with respect to the changes below \tc\ grows rapidly on the underdoped side where it exceeds 50\% at $\tc=35$~K. Notably, similar trends have been previously derived~\cite{Wang06,Li10} and it was pointed out that these cannot be accounted for by Gaussian fluctuations which would involve a much smaller fraction of electronic states but require critical fluctuations that extend over a wide temperature range above \tc~\cite{Emery95,Curty03}. Note that \tons\ remains finite even at $3\%<p<5\%$ where $\tc=0$. This is likely due to quantum fluctuations which suppress the coherency of the condensate~\cite{Emery95b}.

\begin{figure}
\vspace{-1.0cm}
\begin{center}
\hspace*{-0.6cm}\includegraphics[width=10cm]{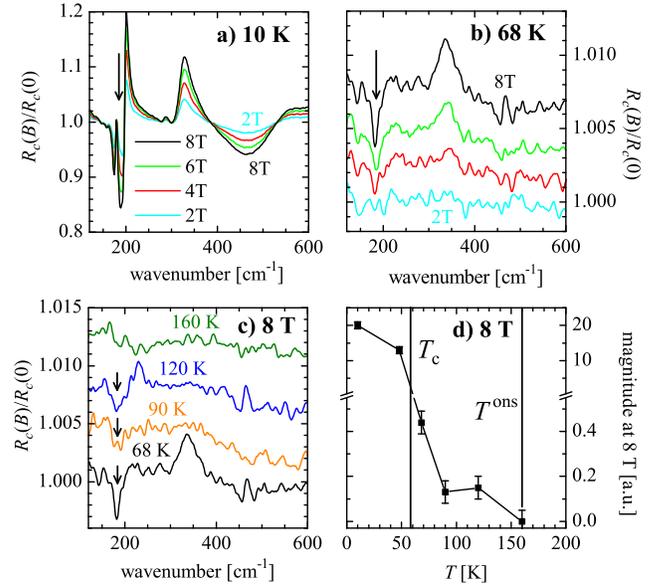} 
\end{center}
\vspace{-1.2cm}
\caption{\label{mag} (a), (b) Ratio of the $c$-axis reflectivity in a magnetic field parallel to the $c$-axis, $R_c(B)$, to the one in zero field, $R_c(0)$, of underdoped Y-123 with $\tc=58$~K. (c) Overview of the 8~T spectra above \tc\ that are shifted for clarity along the vertical scale. Arrows mark the feature around 185~\cm. (d) Temperature dependent magnitude of the feature at 185~\cm. }
\end{figure}

We also find that the phonon anomalies at $\tc<T<\tons$, that are directly related to the TPM, can be partially suppressed with a magnetic field, $B$, similar to the behavior below \tc~\cite{LaForge07}. Figure~\ref{mag} displays representative spectra of the reflectance ratio, $R_c(B)/R_c(0)$, where $B$ is parallel to the $c$-axis of underdoped Y-123 with $\tc=58$~K, $\tons\approx160$~K, and $\tstar>300$~K (for details see~Ref.~\cite{SOM} part II). Figure~\ref{mag}(a) shows the 10~K spectra which demonstrate that the strongest feature is associated with the phonon mode at 185~\cm. 
This is not surprising since this phonon exhibits large temperature dependent changes that are accounted for by the MLM~\cite{Dubroka04} and thus are associated with the intra-bilayer currents. Accordingly, it provides a sensitive indicator to test whether the magnetic field has a similar effect on the coherency of the electronic state at $\tc<T<\tons$ as it has at $T<\tc$. That this is indeed the case is shown in Figs.~\ref{mag}(b)-\ref{mag}(c) where weaker, yet significant and qualitatively similar features appear at 68~K, 90~K and 120~K but are absent at 160~K. We emphasize that this similarity of the magnetic field effects at $T<\tc$ and $\tc<T<\tons$ is the hallmark of a PCS. Figure~\ref{mag}(d) displays a quantification of the anomaly of the 185~\cm\ phonon based on a Lorentz-oscillator-fit of 
the $R_c(8~{\rm T})/R_c(0)$ ratio. The much smaller magnitude of the effect above $\tc$ (than below $\tc$) can be qualitatively understood as follows. Apart from a partial decoupling of the vortex lines, the large field-induced changes of the TPM and the phonon anomalies at $T\ll\tc$ arise because the vortex cores suppress the volume average of the condensate density. This effect is strongly reduced above $\tc$, where a considerable density of spontaneous vortex/antivortex pairs exists already without the magnetic field whose main effect is likely an increase of the net vorticity, i.e. of the difference between the density of vortices and antivortices. The much weaker field effect above $\tc$ is also consistent with the slower suppression at high fields of the magnetization~\cite{Li10}, and the thermal expansion~\cite{Lortz03}.

Next we refer to the corresponding changes in the spin dynamics. The most prominent involves the so-called magnetic resonance mode as measured by inelastic neutron scattering. In optimally-doped Y-123 it emerges right below \tc\ and is recognized as a hallmark of an unconventional d-wave superconductor~\cite{Keimer00}. However, for underdoped Y-123 low-energy magnetic excitations persist well above \tc, where they evolve with temperature in a similar way as the TPM and the phonon anomalies~\cite{Timusk03,Hinkov07}. In addition, on some of the same strongly underdoped Y-123 ($\tc\leq35$~K) crystals it was recently observed that the spin fluctuations develop a characteristic in-plane anisotropy close to our \tons\ which has been interpreted in terms of an electronic transition into a nematic liquid state~\cite{Hinkov08}. Such a transition may give rise to a significant enhancement of the pairing correlations~\cite{Kivelson98} and thus can explain the fact that our infrared data exhibit a relatively sharp anomaly around \tons\ even in the strongly underdoped samples where the condensate density is strongly suppressed. 

In summary, we demonstrated that two different kinds of correlation phenomena lie at the heart of the anomalous normal-state electronic properties of the underdoped cuprate superconductors. One of them is due to a competing pseudogap that gives rise to an insulator-like depletion of the density of low-energy electronic states. The other exhibits signatures of a precursor superconducting state since it enhances the spectral weight of the coherent response and is suppressed by an external magnetic field. Its onset temperature reaches a maximum of $\tons\approx180$~K in the strongly-underdoped regime just before static magnetism develops.

\begin{acknowledgments}
We acknowledge Y.L. Mathis for support at the infrared-beamline of the ANKA synchrotron at KIT, D. The work was financed by the Schweizerischer Nationalfonds (SNF) by grants 200020-119784, 200020-129484,and the NCCR-MaNEP, the Deutsche Forschungsgemeinschaft (DFG) grant BE2684/1 in FOR538, and the Ministry of Education of the CR (MSM002162401). \end{acknowledgments}

\end{document}